\documentclass[aps,twocolumn,showpacs]{revtex4}
 \usepackage{graphicx,amsmath,amssymb}
\usepackage[usenames]{color}
\usepackage[colorlinks=true, urlcolor=navyblue, linkcolor=navyblue, citecolor=navyblue]{hyperref}
\usepackage{relsize}
\definecolor{navyblue}{rgb}{0,0.08,0.45}
\def\Journal#1#2#3#4{{#1} {{\bf #2},} {#3} {(#4)}}

\def\PLB{{Phys. Lett.}  B}

\def\PRL{ Phys. Rev. Lett.}

\def\PRD{{Phys. Rev.} D}

\def\RMP{{Rev. Mod. Phys.}}
\def\MPL{Mod. Phys. Lett.}

\def\JHEP{{J. High Energy Phys.}}
\def\JPG{{J. Phys. G.}}

\def\be{\begin{equation}}
\def\ee{\end{equation}}
\def\bea{\begin{eqnarray}}
\def\eea{\end{eqnarray}}

\def\babar{\mbox{\slshape B\kern-0.1em{\smaller A}\kern-0.1em
    B\kern-0.1em{\smaller A\kern-0.2em R}}}

\begin{document}

\title{Neutrino masses from lepton and quark mass relations and neutrino oscillations}
\author{Fu-Guang Cao}
\affiliation{Institute of Fundamental Sciences, Massey University, Private Bag 11 222, Palmerston North, New Zealand}

\date{\today}

\begin{abstract}

Determining the absolute masses of neutrinos is of fundamental importance in particle physics, nuclear physics, and astrophysics. 
We conjecture that intrinsic mass relations exist between leptons and quarks. Using these relations and neutrino oscillation data,
we show that the inverted neutrino mass hierarchy is strongly disfavored and estimate the absolute neutrino masses to be
$m_{1}=0.21^{+1.70}_{-0.21}  \times 10^{-4}~{\rm eV}$,
$m_{2}=(8.7 \pm 0.1) \times 10^{-3}~{\rm eV}$, and
$m_{3}=(4.9 \pm 0.1) \times 10^{-2}~{\rm eV}$.
\end{abstract}

\maketitle


The study of neutrinos has been at the frontiers of particle physics, nuclear physics, and astrophysics
ever since the particle was first suggested by Fermi and subsequently observed by Cowan and Reines in the 1950s.
Surprising new phenomena related to neutrinos continue revealing the fascinating properties of these
elementary particles and of nature itself \cite{footnote1}.
Neutrinos participating in only the Weak interaction in the Standard Model (SM) and traveling close to the speed of light
make them very difficult to study experimentally.
Neutrino oscillation experiments indicate that neutrinos have mass and thereby the need of extension beyond the SM.
Most models for the extension of the SM suggest that neutrinos could either be Dirac type or Majorana type, depending on whether 
the conservation of lepton number is retained or not.

The absolute masses of neutrinos are an outstanding question puzzling the world of neutrino physics.
The direct measurements of neutrino masses via processes such as neutrinoless double beta decay \cite{NSDBD,SMBilenkyG12} and 
indirect ``measurements" from cosmology  \cite{YWong11} have only been able to impose an upper limit
on the neutrino masses, $\sum^3_{i=1} m_i \lesssim 1~{\rm eV}$.
Many extensions of the SM explain the neutrino masses via the introduction of new heavy particles and/or new energy scales,
and thus have very limited predictive power on the neutrino masses.

We conjecture that close mass relations exist between leptons and quarks after noting that the Koide mass relations \cite{Koide8283}
for the charged leptons and the heavy quarks hold to a surprisingly high degree of accuracy.
This leads to additional constraints on the neutrino masses which, when combined with the neutrino oscillation data,
enables us to determine the three absolute neutrino masses.

The charged lepton masses obey the empirical Koide relation \cite{Koide8283},
\bea
K^l_{heavy}=\frac{m_{e}+m_{\mu}+m_{\tau}}{\left(\sqrt{m_{e}}+\sqrt{m_{\mu}}+\sqrt{m_{\tau}}\right)^2} \simeq \frac{2}{3},
\label{eq:Koide}
\eea
with surprisingly high precision.
Using the latest values for the charged lepton masses given by the Particle Data Group 2010 (PDG 2010) \cite{PDG2010}
we have $K=0.666658^{+0.000009}_{-0.000015}$.
It has long been advocated \cite{Koide90s,Foot94,Krolikowski05,RiveroG05,GerardGH06,Sumino} that grand-unified extensions of the SM
might lead to intriguing relations such as Eq.~(\ref{eq:Koide}) for the neutrinos and quarks \cite{NLiM05,Kartavtsev11,RZhang11}.

Recently, Kartavtsev  \cite{Kartavtsev11} and Rodejohann and Zhang \cite{RZhang11} proposed a generalization of the Koide relation to the quark sector
and defined three Koide-like parameters,
\bea
K^q_{light}&=& \frac{m_u+m_d+m_s}{\left(\sqrt{m_u}+\sqrt{m_d}+\sqrt{m_s}\right)^2}, \label{eq:Kql} \\
K^q_{heavy}&=& \frac{m_c+m_b+m_t}{\left(\sqrt{m_c}+\sqrt{m_b}+\sqrt{m_t}\right)^2}, \label{eq:Kqh} \\
K_{quark}&=& \frac{\sum_i m_{q_i}}{ \left( \sum_i \sqrt{m_{q_i}} \right)^2 }. \label{eq:Kq}
\eea
It was found \cite{Kartavtsev11} that $K^q_{heavy}$ and $K_{quark}$ are closer to the Koide limit of $2/3$ than $K^q_{light}$,
though dividing the quarks into the light and the heavy ones puts the $s$ and $c$ quarks belonging to the second generation of
fermions in the SM into different groups.
Using the quark mass values given by the Particle Data Group 2010~\cite{PDG2010} 
we find
\bea
K^q_{light} &=&0.566^{+0.039}_{-0.030}, \label{eq:KqlNum}\\
K^q_{heavy}&=&0.6688^{+0.0034}_{-0.0038}, \label{eq:KqhNum}\\
K_{quark}&=&0.6349^{+0.0048}_{-0.0042}. \label{eq:KqNum}
\eea
The Koide parameters defined for the charged lepton and heavy quarks, $K^l_{heavy}$ and $K^q_{heavy}$, 
are in excellent agreement (differing by less than $0.3\%$),
which is quite remarkable since the masses involved are vastly different, varying from about $0.5$~MeV to about $170$~GeV. 

Rodejohann and Zhang assumed that the Koide relation exists in the Dirac and Majorana mass matrices
in the seesaw mechanism for the generation of neutrino masses after noting that the exact Koide relation may not hold for neutrinos \cite{RZhang11}.

We propose that Koide-Kartavtsev-Rodejohann-Zhang factors defined in a similar way to Eqs.~(\ref{eq:Kql}), (\ref{eq:Kqh}), and (\ref{eq:Kq}) for the lepton sector
have the same values as for the quark sector, 
\bea
K^l_{light}&=& \frac{m_{1}+m_{2}+m_{3}}{\left(\sqrt{m_{1}}+\sqrt{m_{2}}+\sqrt{m_{3}}\right)^2} \nonumber \\
&\simeq& K^q_{light}, \label{eq:Kll} \\
K^l_{heavy}&=& \frac{m_{e}+m_{\mu}+m_{\tau}}{\left(\sqrt{m_{e}}+\sqrt{m_{\mu}}+\sqrt{m_{\tau}}\right)^2} \nonumber \\
&\simeq& K^q_{heavy}, \label{eq:Klh} \\
K_{lepton}&=& \frac{m_{e}+m_{\mu}+m_{\tau}+\sum_{i=1}^3 m_i}{\left( \sqrt{m_{e}} + \sqrt{m_{\mu}}+\sqrt{m_{\tau}}+\sum_{i=1}^3 \sqrt{m_i}\right)^2}. \nonumber \\
&\simeq& K_{quark} \label{eq:Kl} 
\eea
Equations (\ref{eq:Kll}) and (\ref{eq:Kl}) provide new constraints on the neutrino masses.
However, the value of $K_{lepton}$ is dominantly decided by the masses of the charged leptons
unless the neutrino masses are of the order of MeV, a very unlikely situation when we consider the constraints
from neutrinoless double beta decay measurements \cite{NSDBD} and cosmological observations \cite{YWong11}.
Thus Eq.~(\ref{eq:Kl}) does not put a strong constraint on the neutrino masses.

Equation (\ref{eq:Klh}) is satisfied with a very high degree of accuracy
while $K_{lepton}$ and $K_{quark}$ differ by less than $5\%$ when the neutrino mass terms are ignored.
We may expect that the relation $K^l_{light}\simeq K^q_{light}$ is respected at higher degrees of accuracy than $K_{lepton}\simeq K_{quark}$;
thus, Eq.~(\ref{eq:Kll}) provides a strong constraint on the neutrino masses, {\it i.e.},
\bea
\frac{m_{1}+m_{2}+m_{3}}{\left(\sqrt{m_{1}}+\sqrt{m_{2}}+\sqrt{m_{3}}\right)^2}=0.566^{+0.039}_{-0.030}.
\label{eq:constraint3}
\eea

Neutrino oscillations are sensitive to neutrino mass squared difference and
the latest measurements suggested~\cite{SchwetzTJV},
\bea
\Delta m_{21}^2 =m_{2}^2-m_{1}^2=7.65 ^{+0.23}_{-0.20} \times 10^{-5}~{\rm eV}^2, \label{eq:mixing1}\\
\left| \Delta m_{31}^2 \right| = \left| m_{3}^2-m_{1}^2 \right| =2.40 ^{+0.12}_{-0.11} \times 10^{-3}~{\rm eV}^2. \label{eq:mixing2}
\eea
Two possible neutrino mass hierarchies compatible with the neutrino oscillation data are
the normal mass hierarchy, $m_{1} < m_{2} \ll m_{3}$, and the inverted mass hierarchy, $m_{3} \ll m_{1} < m_{2}$.

\begin{figure}[htbp]
\begin{center}
\includegraphics[width=7.5cm]{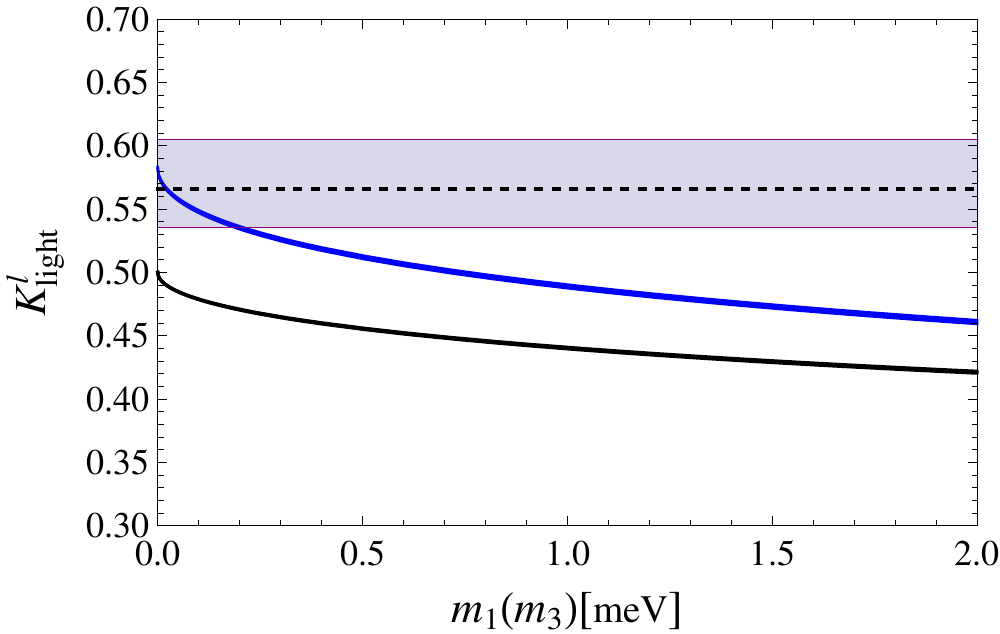}
\caption{The dependence of $K^l_{light}$ on mass of the lightest neutrino ($m_{1}$  or $m_{3}$) for the normal mass hierarchy (upper blue curve)
and inverted mass hierarchy (lower black curve). The band of each curve represents the uncertainty allowed by
varying $\Delta m^2_{21}$ and $\left| \Delta m^2_{31} \right|$ in their $1\sigma$ intervals.
The horizontal band represents the allowed $1\sigma$ range of $K^q_{light}$.}
\label{fig:Kllight}
\end{center}
\end{figure}

Substituting Eqs.~(\ref{eq:mixing1}) and (\ref{eq:mixing2}) into the first line of Eq.~(\ref{eq:Kll}), we can study the dependence of $K^l_{light}$ 
on the mass of the lightest neutrino, $m_{1}$ for the normal mass hierarchy or $m_{3}$ for the inverted mass hierarchy.
The results are shown in Fig.~\ref{fig:Kllight}.
We can see that $K^l_{light}$ approaches the maximum value when the mass of the lightest neutrino goes to zero for both cases of 
neutrino mass hierarchy, and the difference between $K^l_{light}$ and $K^q_{light}$ increases with $m_{1}$ or
$m_{3}$ increasing. Thus, Eq.~(\ref{eq:Kll}) favors a small value for the mass of the lightest neutrino.
In the case of an inverted mass hierarchy, the maximum value of $K^l_{light}$ is smaller than 
$K^q_{light}$ by $2\sigma$,
which indicates that the inverted mass hierarchy is not compatible with our conjecture of the same Koide-like mass relation
for neutrinos and the light quarks.

Equation~(\ref{eq:Kll}) could be satisfied in the case of a normal mass hierarchy.
We find the allowed range of $m_{1}$ by varying $K^q_{light}$ in its $1\sigma$ interval and the result is 
$0~{\rm eV} \lesssim m_{1} \lesssim 1.9 \times 10^{-4}$ eV with a central value of $2.1 \times 10^{-5}~{\rm eV}$.
Thus we obtain \cite{footnote2}
\bea
m_{1}&=&0.21^{+1.70}_{-0.21}  \times 10^{-4}~{\rm eV}, \nonumber \\
m_{2}&=&(8.7 \pm 0.1) \times 10^{-3}~{\rm eV}, \label{eq:numasses} \\
m_{3}&=&(4.9 \pm 0.1) \times 10^{-2}~{\rm eV}. \nonumber
\eea
The values of $m_{2}$ and $m_{3}$ are predominately decided by the neutrino oscillation data since
 $m_{1} \ll m_{2} \ll m_{3}$.
To measure such small neutrino masses as we reported here poses a significant challenge to the current and upcoming experiments.
  
The neutrinoless double-beta decay process is sensitive to the effective Majorana mass defined as \cite{NSDBD,SMBilenkyG12}
\bea
m_{\beta \beta}= \left| \sum_{i=1}^3 V^2_{e i} m_i \right|,
\eea
where $V_{e i}$ are elements of the leptonic flavor mixing matrix which can be expressed in terms of neutrino mixing angles, and CP-violating phases.
Using the latest experimental measurements for the neutrino mixing angles, $\sin^2 \theta_{12}=0.304$, $\sin^2 \theta_{23}=0.5$, \cite{PDG2010} and
$\sin^2 2 \theta_{13}=0.092$ \cite{DayaBay}, and under the assumption of CP invariance
we obtain $m_{\beta \beta}=0.0039$~eV. The value is well below the current upper bound imposed by the neutrinoless double-beta decay experiments,
$m_{\beta \beta} \lesssim (0.22 - 1.00 )$ eV, but is within the range of sensitivity of several experiments in preparation,
$m_{\beta \beta} \lesssim 0.01$~eV \cite{SMBilenkyG12}. 
The sum of neutrino masses evaluated with Eq.~(\ref{eq:numasses}), $\sum_{i=1}^3 m_i=0.058$ eV, is consistent with the upper limit obtained from the probes of the 
large-scale structure of the universe and of the cosmic microwave background anisotropies, $\sum_{i=1}^3 m_i \lesssim 1$~eV \cite{YWong11}. 

The neutrino masses are determined by the masses of much heavier Dirac and Majorana particles in the seesaw mechanism for the neutrino masses,
\bea
m_i=M_D M_R^{-1} M_D^T,  \label{eq:seesaw}
\eea
where $M_D$ and $M_R$ are Dirac and right-handed Majorana mass terms, respectively.
One may evaluate the masses of these new heavy particles using the absolute masses obtained in Eq.~(\ref{eq:numasses}).
For example, using a simple model studied in \cite{RZhang11} which assumed both the Dirac and Majorana mass matrixes are diagonal,
{\it i.e.}, $M_D={\rm diag}(D_1,D_2,D_3)$ and $M_R={\rm diag}(M_1,M_2,M_3)$,
we find the three heavy Dirac neutrino masses are $D_1\sim (0.46 \sim 1.38) \times 10^2$ GeV, $D_2\sim (9 \sim 30) \times 10^2$ GeV, 
and $D_1\sim (2.2 \sim 6.0) \times 10^3$ GeV for $M_i \sim 10^{14\sim 15}$ GeV being just below the typical scale of grand-unified theories
$\Lambda_{\rm GUT} \sim 10^{16}$~GeV. It may be possible to detect these heavy Dirac neutrinos at the Large Hadron Collider \cite{HanZ06}.

In summary, determining the absolute masses of neutrinos is of fundamental importance in understanding the Standard Model 
and new physics beyond the Standard Model.
We proposed intrinsic mass relations for leptons and quarks that lead to a strong constraint on the neutrino masses.
Combining this constraint with the neutrino oscillation data we showed that the inverted neutrino mass hierarchy is strongly disfavored
and we determined the absolute neutrino masses for the normal mass hierarchy.

I would like to thank J. Y. Lu and Y. G. Cao for carefully reading the manuscript.


\begin{thebibliography}{00}
\bibitem{footnote1}
The recent observation of superluminal neutrinos reported by the OPERA Collaboration \cite{OPERA}
provides such an example. If confirmed, the OPERA experiment could indicate new physics beyond the Standard Model.
\bibitem{footnote2}
Allowing Eq.~(\ref{eq:Kll}) to be violated by $5\%$ 
(the same level of possible violation of $K_{lepton} \simeq K_{quark}$ when the neutrino mass terms are ignored),
and using the running quark masses at the scale of $M_Z=91.2$~GeV~\cite{ZXingZZ08}
will not change the conclusions drawn above.
\end{thebibliography}
\end{document}